\pdfoutput=1
\documentclass[a4paper,fleqn,usenatbib]{mnras}
\usepackage{graphicx}	
\usepackage{amsmath}	
\usepackage{amssymb}	
\usepackage{bm}
\usepackage{txfonts}

\usepackage[T1]{fontenc}
\usepackage{ae,aecompl}

\title[Interaction of eccentric SMBBH with intermediate mass ratio and circumbinary accretion disk]{Interaction of eccentric supermassive binary black hole with intermediate mass ratio and circumbinary accretion disk}

\author[Wenshuai Liu]{
Wenshuai Liu\thanks{E-mail: 674602871@qq.com}\\
School of Physics, Henan Normal University, Xinxiang 453007, China\\
}

\date{Accepted XXX. Received YYY; in original form ZZZ}

\begin{document}
\label{firstpage}
\pagerange{\pageref{firstpage}--\pageref{lastpage}}
\maketitle

\begin{abstract}
Recent simulations show that the eccentricity of supermassive binary black hole with intermediate mass ratio could grow toward near unity through gravitational interaction with the stellar background in the merging remnant after two galaxies merge. The increased eccentricity reduces the timescale of the supermassive binary black hole merger through the strong gravitational radiation at periastron. Usually, large amount of gas flows toward the center of the newly merged galaxy, forming circumbinary gaseous disk around the binary in the center of the newly merged galaxy. Tidal interaction between such eccentric binary with intermediate mass ratio and circumbinary disk need to be investigated quantitatively. In this work, we study the gravitational interaction of the eccentric supermassive binary black hole with intermediate mass ratio and the circumbinary disk using code FARGO3D. Simulations are carried out with different semimajor, eccentricity and mass ratio. We find that the accretion rate onto the inner boundary could be strongly affected by the secondary black hole and tend to present periodic accretion rate in some situations. Such periodic accretion rate can be used as electromagnetic counterpart to the gravitational wave radiated by such kind of eccentric binary.

  \end{abstract}

\begin{keywords}
accretion, accretion discs -- black hole physics -- gravitational waves -- hydrodynamics -- methods: numerical
\end{keywords}



\section{Introduction}

Supermassive black holes (SMBH) with mass of $10^6$M$_{\odot}$-$10^9$M$_{\odot}$ residing in the center of galaxies \citep{Kormendy95,Ferrarese05,Kormendy13} are considered to co-evolve with their host galaxies, leading to the bound supermassive binary black hole (SMBBH) formed from galactic merger according to the hierarchical picture of structure formation \citep{White78}. When two galaxies collide, the dynamical friction of the field stars and the two SMBHs in the merged galaxy will lead to the shrink of the semimajor axis of the SMBBH \citep{Milosavljevic01}. After the ejection of all of the nearby stars by SMBBH, the semimajor axis will stall without the possibility that they merge within the Hubble time if only considering gravitational radiation as the shrink mechanism afterward. This is usually called the "last-parsec problem" due to the fact that the typical stalled separation of the SMBBH is about $\sim$pc, which is confirmed by many researches with N-body simulations \citep{Makino04,Berczik05}. To overcome this bottleneck, a possible mechanism to extract energy and angular momentum from SMBBH is proposed that the dynamical interaction with the surrounding gaseous disk \citep{Armitage05,MacFadyen08,Cuadra09,Lodato09} will be able to reduce the semimajor axis of the SMBBH down to the stage where gravitational radiation will act as the dominant mechanism for the reduction of the SMBBH separation. Gravitational wave radiated by SMBBH will be the target for proposed spacebased laser interferometers \citep{Schnittman11} and the ongoing International Pulsar Timing Array \citep{Hobbs10}. Electromagnetic counterparts to gravitational wave radiated by massive binary black hole (MBBH) has been paid much attention with respect to coalescing MBBH which will be observed by LISA \citep{Schnittman11} with the prediction that possible counterparts of violent shock in the gas disk surrounding the final coalescence may exist. The majority of the targets of LISA is clustered at $3<z<8$ with frequency range from $10^{-4}$ to $10^{-1}$ Hz \citep{Sesana07}, and such targets are expected to be MBBH with mass of $\leq 10^6 M_\odot$ \citep{Sesana07}. Targets of PTA, emitting in the frequency range of $10^{-9}$ to $10^{-6}$ Hz, are those with mass $\geq 10^8 M_\odot$ and redshift $z<1$ \citep{Sesana09,Lee11}, making electromagnetic identification of the gravitational wave source easier than that by LISA.

Most of previous studies on the evolution of the SMBBH have mainly focused on the evolution of the semimajor axis of the SMBBH. Due to the large amount of gravitational radiation at periastron from SMBBH with large eccentricity, eccentricity of SMBBH is expected to play an important role in reducing the time for coalescence at a given semimajor axis of the SMBBH. Shrinking mechanisms based on gravitational interaction with both stellar and gas show that the eccentricity of the SMBBH can be increased efficiently. Results from \cite{Iwasawa11} show that the interaction of stellar and the SMBBH with intermediate mass ratio can excite the eccentricity of the SMBBH to be near unity, indicating that the coalescence timescale associated with gravitational radiation can be much shorter for SMBBH with large eccentricity. Similar results of increase of the
eccentricity by interaction of stellar and SMBBH at sub-pc scales are shown in \cite{Sesana10,Sesana11}. Eccentricity excitation due to interaction of SMBBH and gaseous disk studied in \cite{Hayasaki09} shows that eccentricity of SMBBH can reach up to near unity along with the evolution of semimajor axis of the SMBBH.

Previous works \citep{Hayasaki07,Hayasaki08,Hayasaki13,Roedig11} have studied SMBBH-disk interaction where the mass ratio of SMBBH is mild and eccentricity is relatively small. Thus, it's an important goal to study the interaction of SMBBH with large eccentricity and intermediate mass ratio and circumbinary disk. In this work, we investigate the gravitational interaction of SMBBH with intermediate mass ratio and large eccentricity and the gaseous disk they are embedded in. The results show that the accretion rate onto the inner boundary affected by the secondary black hole depends on the eccentricity, mass ratio and semimajor of the SMBBH during the evolution, which may provide significant electromagnetic feature along with the gravitational wave radiated by such SMBBH..

This work is organized as follows. In Section 2, we describe the physical problem and the initial condition of simulations of the SMBBH-disk interaction. In Section 3, we present the results from the hydrodynamical simulations and analyses from the point of view of potential observational features. Conclusions are given in Section 4.

\section{PHYSICAL PROBLEM and INITIAL CONDITION}
We study the gravitational interaction between an eccentric SMBBH with an intermediate mass ratio at sub-pc scales and the circumbinary disk it is embedded in after the dynamical process of interaction with the stellar and gaseous environment. Emission of gravitational waves will become the dominant mechanism to extract the binary's orbital energy when the binary's separation is small enough. In this study, we define the mass ratio to be $q=\frac{M_2}{M_1}$ where $M_1$ and $M_2$ are the mass of the primary black hole and the secondary black hole, respectively. The evolution equations for the semimajor and eccentricity of the SMBBH are given by \citep{Peters63}.

\begin{equation}
\frac{da}{dt}\big|_{\rm gw}=-\frac{64}{5}\frac{G^3}{c^5}\frac{M_1M_2M}{a^3(1-e^2)^{7/2}}\left(1+\frac{73}{24}e^2+\frac{37}{96}e^4\right)
\label{a}
\end{equation}

\begin{equation}
\frac{de}{dt}\big|_{\rm gw}=-\frac{304}{15}\frac{G^3}{c^5}\frac{M_1M_2M}{a^4(1-e^2)^{5/2}}e\left(1+\frac{121}{304}e^2\right)
\label{e}
\end{equation}

The corresponding merging timescale of the binary due to gravitational radiation is

\begin{equation}
\frac{\tau_{\rm{GW}}}{P_{\rm{orb}}}\sim6\times10^{5}\frac{(1+q)^{2}}{q}f(e)\left(\frac{0.01\rm{pc}}{a}\right)^{-5/2}\left(\frac{M_{\rm{BH}}}{10^{8}M_{\odot}}\right)^{-5/2} \label{1}
\end{equation}
where
\begin{equation}
f(e)=(1-e^2)^{7/2}/(1+73e^{2}/24+37e^{4}/96)
\end{equation}
and $P_{\rm{orb}}$ is the orbital period.

The evolution rate of semimajor axis and eccentricity of SMBBH strongly depend on $a$ and $e$, and decreases to zero with initial small separation and large eccentricity effectively. With such orbital parameters, the secondary black will experience orbital precession due to the effect from general relativity. Thus, we adopt the first order post-Newtonian orbital equation for the secondary black hole around the primary as follows

\begin{equation}
\frac{d \bm{v}}{dt}=-\frac{Gm}{r^2}[(1-A_{1PN})\bm{n} - \frac{\bm{n} \cdot \bm{v}}{c}B_{1PN}\frac{\bm{v}}{c}]
\label{n}
\end{equation}
where we define $\bm{x}=\bm{x_1}-\bm{x_2}$, $\bm{x_1}$ and $\bm{x_2}$ are the positional vectors of primary and secondary black hole relative to the center of mass of the binary, respectively. $r=|{\bf x}|$, $n^i={x^i}/{r}$, $\nu = M_1M_2/(M_1+M_2)^2$, $m=M_1+M_2$
and
\begin{align}
A_{1PN} = & \frac{3}{2}\nu\left(\frac{\bm{n} \cdot \bm{v}}{c}\right)^{2} -
                                (1+3\nu)\frac{v^{2}}{c^{2}}  + \left(4+2\nu\right)\frac{Gm}{r}\,,\\[1mm]
B_{1PN} = & 4-2\nu\
\end{align}

\begin{table}
 \centering
 \begin{tabular}{l | c | c | c | c}
 \hline
 SIMULATION & $M_1$ & $q$ & $a_0$ & $e$\\
 \hline
 A1 & $5\times10^{8}M_\odot$ & 0.02 & 11.8$R_s$ & 0.6   \\ 
 A2 & $5\times10^{8}M_\odot$ & 0.02 & 11.8$R_s$ & 0.7   \\ 
 B1 & $5\times10^{8}M_\odot$ & 0.01 & 11.8$R_s$ & 0.6  \\ 
 B2 & $5\times10^{8}M_\odot$ & 0.01 & 11.8$R_s$ & 0.7   \\ 
 C1 & $5\times10^{8}M_\odot$ & 0.005 & 11.8$R_s$ & 0.7   \\ 
 D1 & $5\times10^{8}M_\odot$ & 0.01 & 23.6$R_s$ & 0.85   \\ 
 D2 & $5\times10^{8}M_\odot$ & 0.02 & 23.6$R_s$ & 0.85   \\ 
 \hline
 \end{tabular}
\caption{Summary of the simulations performed, from left to right are the name of the simulation, the mass of the primary black hole, the mass ratio of the binary, the semimajor of the binary where $r_g=\frac{2GM_1}{c^2}$, the eccentricity of the binary.}
\label{tab:simoverview}
\end{table}

\begin{figure*}
   \begin{center}
     \begin{tabular}{cc}
     \includegraphics[width=0.33\textwidth]{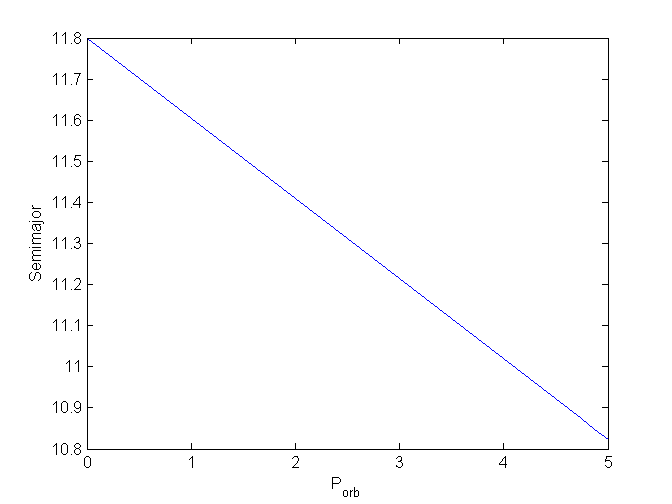}
     \includegraphics[width=0.33\textwidth]{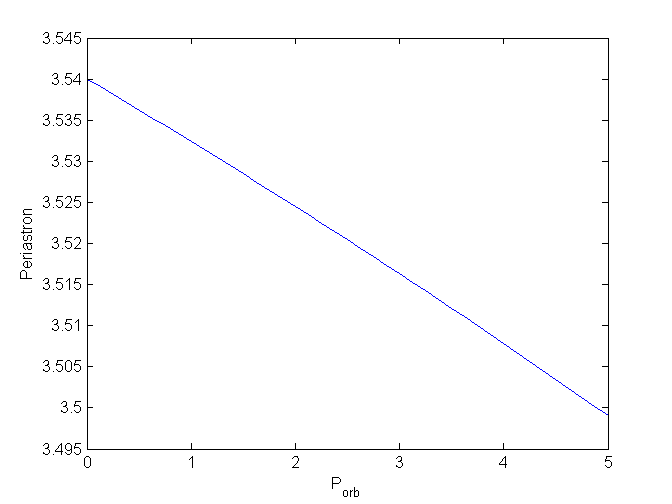}
     \includegraphics[width=0.33\textwidth]{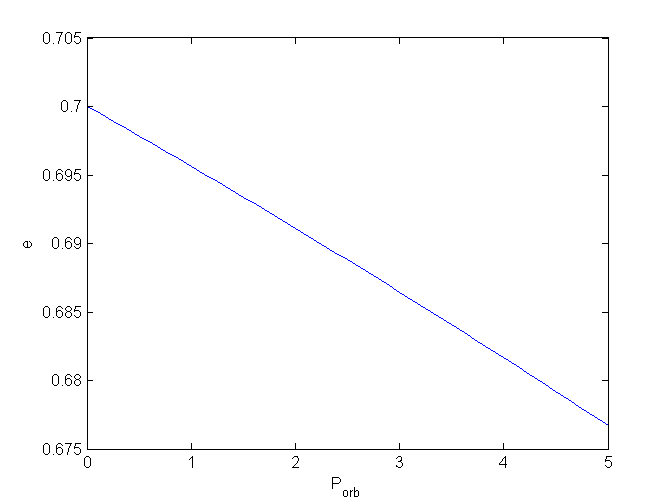}\\
            \end{tabular}
   \end{center}

    \caption{Evolution of semimajor axis (left panel), periastron (middle panel) and eccentricity (right panel) within five orbits with the initial parameters in A2 using equation (\ref{a}) and (\ref{e}).}
    \label{fig:figure6}
\end{figure*}

\begin{figure*}
   \begin{center}
     \begin{tabular}{cc}
     \includegraphics[width=0.5\textwidth]{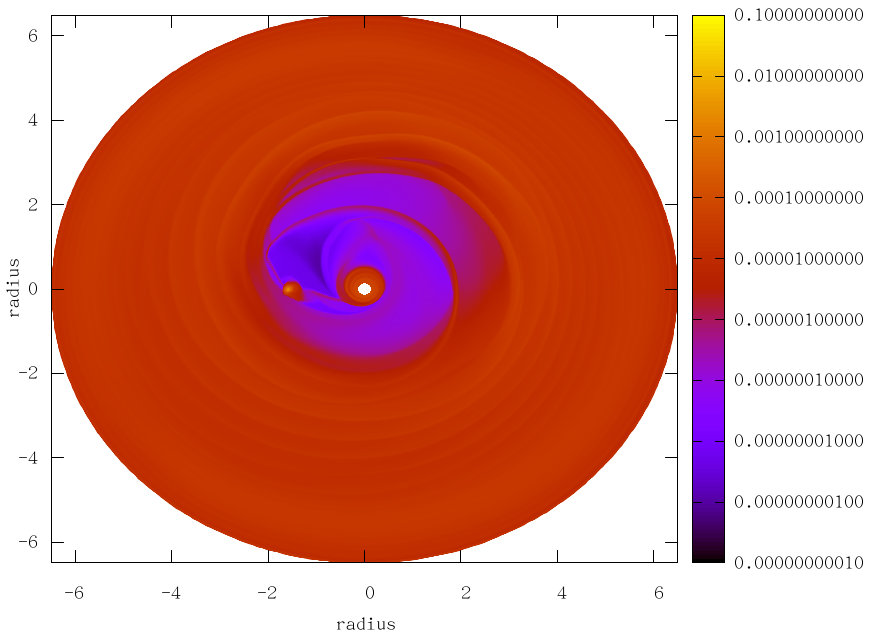}
     \includegraphics[width=0.5\textwidth]{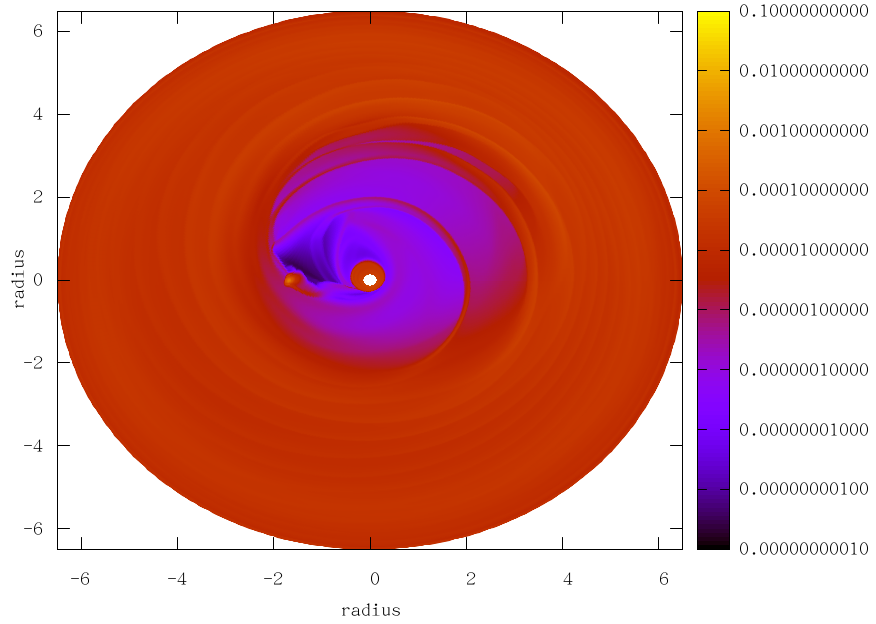}\\
     \includegraphics[width=0.5\textwidth]{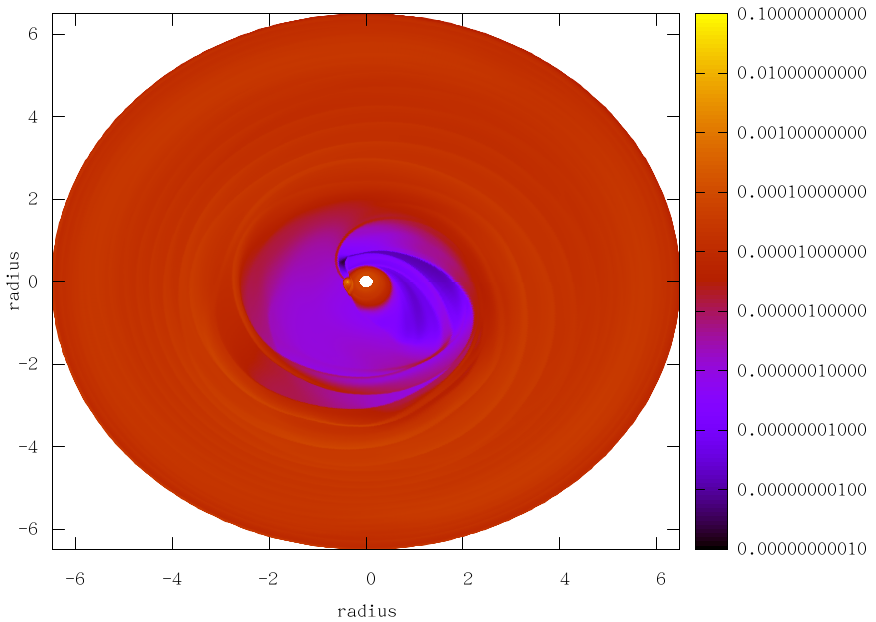}
     \includegraphics[width=0.5\textwidth]{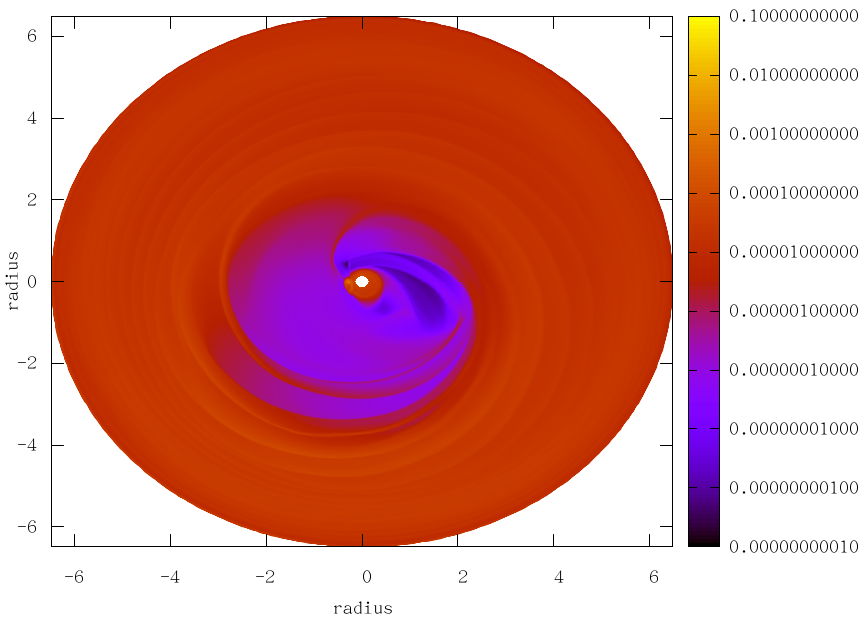}
            \end{tabular}
   \end{center}

    \caption{Up and bottom show the surface density at apastron and periastron after about 400 orbits of the secondary black hole around the primary one, respectively. From left to right, the eccentricity of the secondary black hole is 0.6 and 0.7, respectively.}
    \label{fig:figure1}
\end{figure*}

Eccentricity excitation of SMBBH by interaction with the stellar studied in \cite{Iwasawa11} shows that the eccentricity can reach near 0.999 with the semimajor of about 4 pc from the simulation with primary SMBH of mass $10^{10}M_\odot$ and mass ratio of 0.01, meaning that the periastron can be near $4R_s$ of the primary SMBH. At such small periastron, the accretion onto around the primary will be strongly affected by the secondary while the gravitational radiation is the largest at periastron, thus, gravitational radiation may have electromagnetic counterpart if gaseous disk exists in such situation. In this work, we set the eccentricity of the SMBBH embedded in the disk with relatively large value and the periastron outside the innermost stable circular orbit. Large semimajor with the same periastron will be much time expensive in hydrodynamical simulation due to the relatively higher resolution in order to simulate the accretion onto the primary black hole.

During the tidal interaction with the circumbinary disk, circum-primary and circum-secondary disk will form. Whether disk formed around individual black hole could long live rely on the ratio of the viscous timescale of the disk of size $R_d$ around individual black hole to the orbital period of the binary with the assumption that the disk is geometrically thin and isothermal with the Shakura-Sunyaev viscosity parameter $\alpha_{SS}$. The ratio is \citep{Hayasaki08}

\begin{equation}
\frac{\tau_{\rm{vis}}}{P_{\rm{orb}}}
\sim
5.1\times10^{5}\frac{1}{\sqrt{1+q}}\left(\frac{R_{\rm{d}}}{a}\right)^{1/2}
\left(\frac{0.1}{\alpha_{\rm{SS}}}\right)\left(\frac{10^{4}K}{T}\right)\left(\frac{0.01\,\rm{pc}}{a}\right)\left(\frac{M_{\rm{BH}}}{10^{8}M_{\odot}}\right).
\label{2}
\end{equation}

We conduct several simulations with different mass of SMBBH and eccentricity. The details are shown in Table \ref{tab:simoverview}. With parameters given in Table \ref{tab:simoverview}, it is shown from equation (\ref{1}) that the timescale for merger is several hundred times the binary orbital period. From equation (\ref{2}), the viscous timescale is much longer than the orbital period of the SMBBH, meaning that accretion disk formed around individual black hole is persistent during the evolution of the eccentric SMBBH.

Figure.~\ref{fig:figure6} shows the evolution of semimajor axis and eccentricity within five orbits with the parameters in A2 by integrating equation (\ref{a}) and (\ref{e}). It is clear seen that the change of semimajor axis and eccentricity within the five orbits is minor. So, we study the accretion rate onto the primary black hole by assuming that the semimajor axis and eccentricity of the SMBBH is fixed within five orbit for simplicity. Firstly, we simulate the interaction of the SMBBH and the disk up to 400 orbits of the secondary around the primary so as to the disk attains a quasi-steady state density profile with equation (\ref{n}). Then we simulate the accretion rate onto the primary by restarting the simulation.

We carry out two-dimensional non-selfgravitating hydrodynamical simulations of viscous accretion disks around the primary SMBH which is orbited by the secondary black hole using code FARGO3D \citep{Ben16,Masset00} to study the effect of the eccentric secondary black hole on the accretion onto the primary black hole with cylindrical coordinate system centered onto the primary SMBH. For simplicity, we neglect the force exerting on the secondary black hole by the disk. We set $G=1$, $M_1=1$, $a_0=1$ in code unit. The initial surface density of the accretion disk is $\Sigma=3.1 \times 10^{-5}(\frac{r}{a})^{-0.5}$ extending from $1.77r_g$ to $76.7r_g$ (from 0.15 to 6.5 in code unit) in A1 to C1 and from $1.88r_g$ to $141.6r_g$ (from 0.08 to 6 in code unit) in D1 and D2 with aspect ratio to be 0.05 and the uniform kinematic viscosity of the disk to be $10^{-5}$. The mesh has $N_r=768$ cells along the radial direction so as to there is at least two cell between the inner boundary and the boundary of the Hill sphere of the secondary black hole along the radial direction when approaching periastron and $N_\phi=384$ spanning in azimuth [-$\pi$,$\pi$].


\section{Results and Analyses} \label{discuss}

\begin{figure}

     \includegraphics[width=0.5\textwidth]{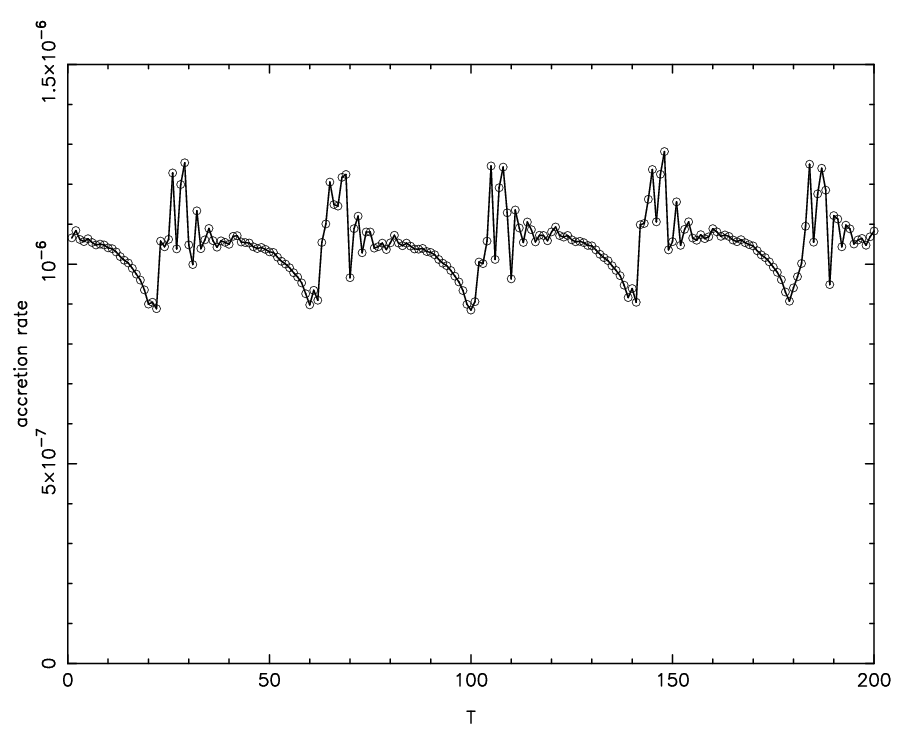}
     \includegraphics[width=0.5\textwidth]{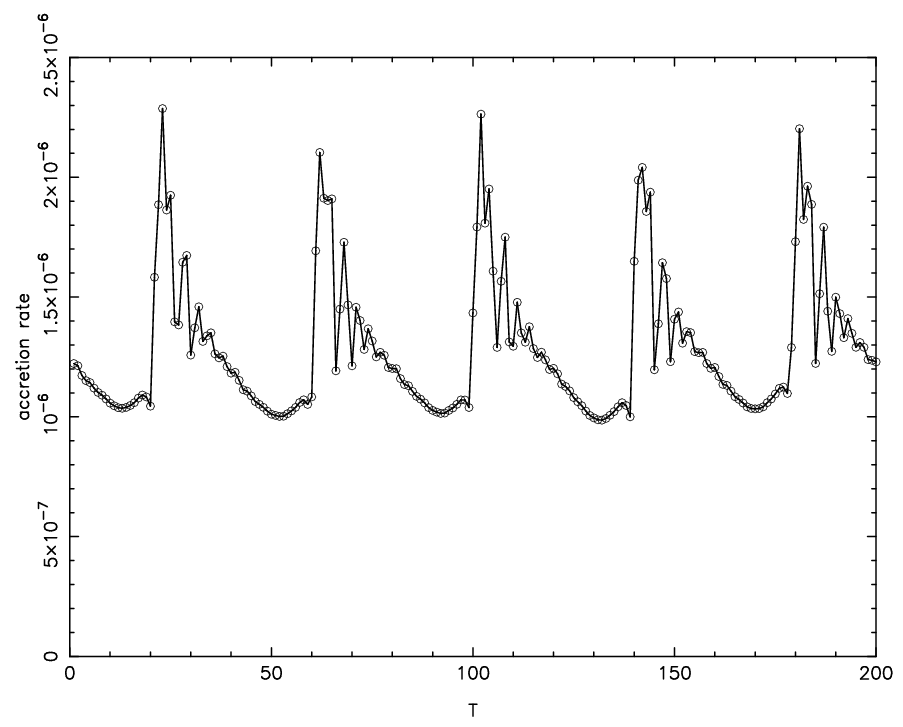}
     \includegraphics[width=0.5\textwidth]{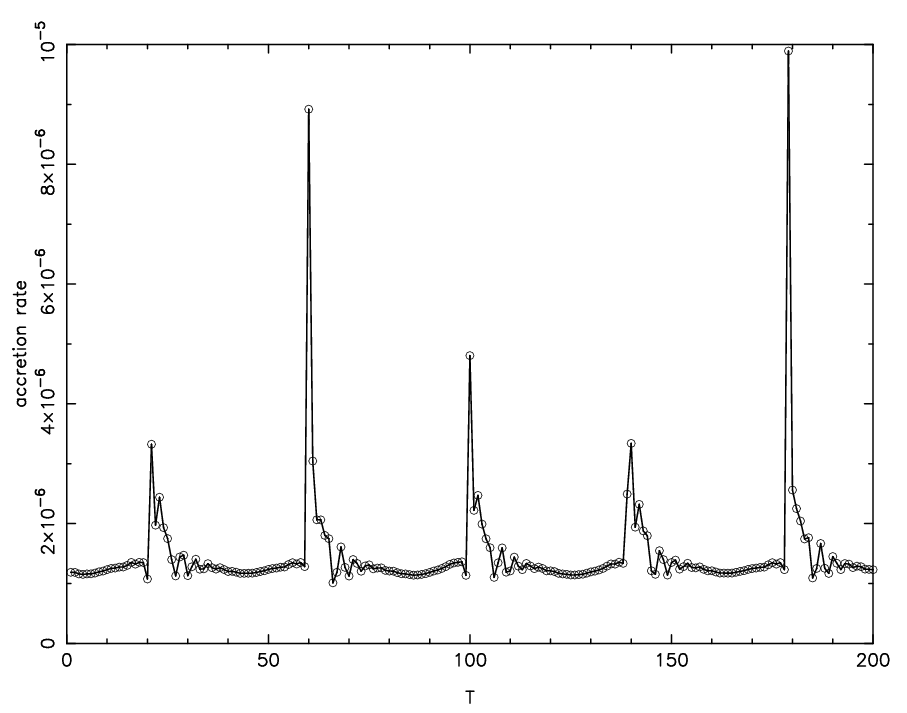}

    \caption{The accretion rate onto the inner boundary within five orbits of the secondary around the primary SMBH after about 400 orbits with eccentricity of 0.6 and 0.7 from up to middle with parameters in A1 and A2. The bottom shows the result with eccentricity of 0.8. The horizontal axis is in unit of $P_{orb}/40$ where $P_{orb}$ is the orbital period of the secondary around the primary and the vertical axis is in code unit.}
    \label{fig:figure2}
\end{figure}

\begin{figure}

     \includegraphics[width=\columnwidth]{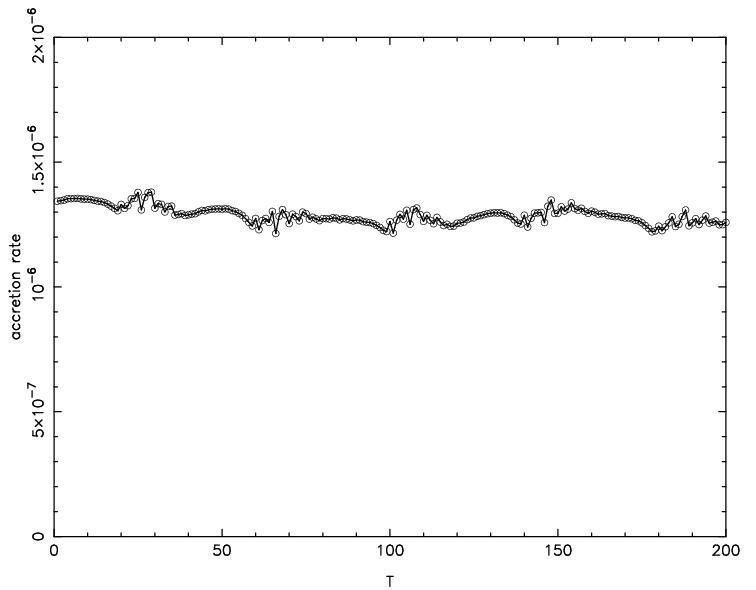}
     \includegraphics[width=\columnwidth]{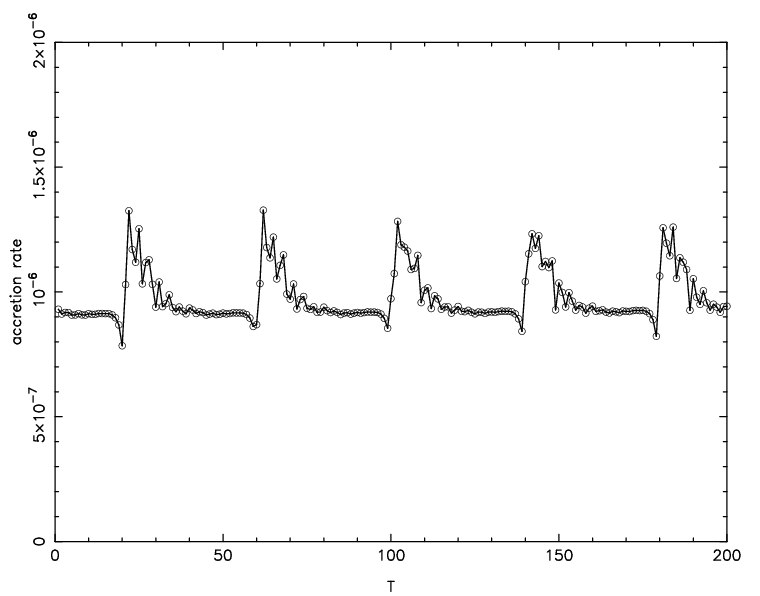}

    \caption{Same as Figure.~\ref{fig:figure2} but for B1 (up panel) and B2 (bottom panel).}
    \label{fig:figure3}
\end{figure}

\begin{figure}

     \includegraphics[width=\columnwidth]{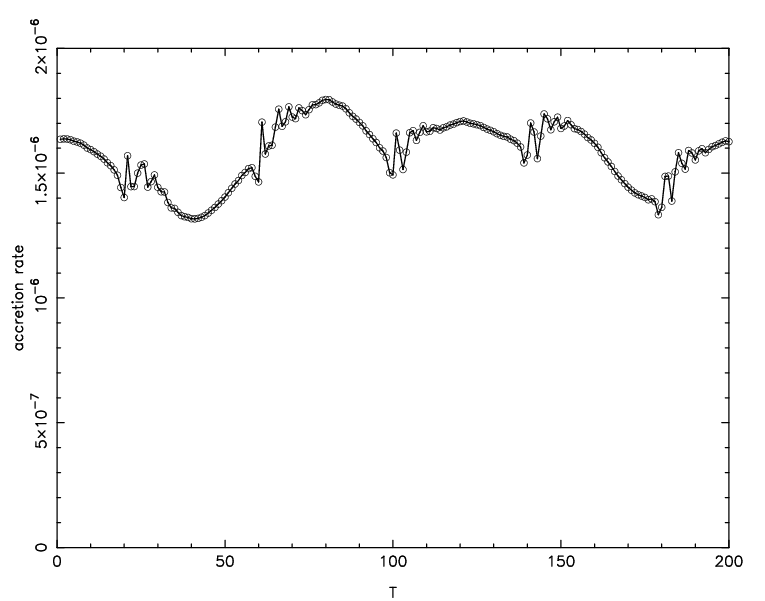}

    \caption{Same as Figure.~\ref{fig:figure2} but for C1.}
    \label{fig:figure4}
\end{figure}

\begin{figure}

     \includegraphics[width=\columnwidth]{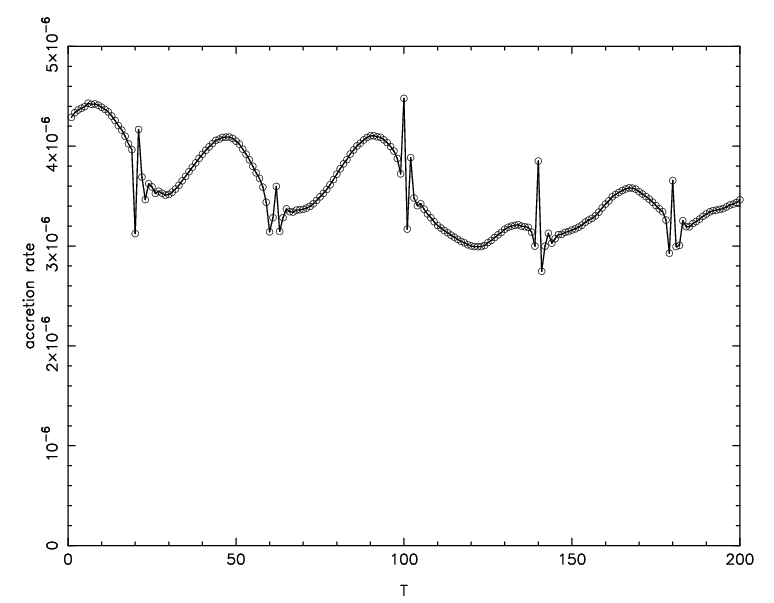}
     \includegraphics[width=\columnwidth]{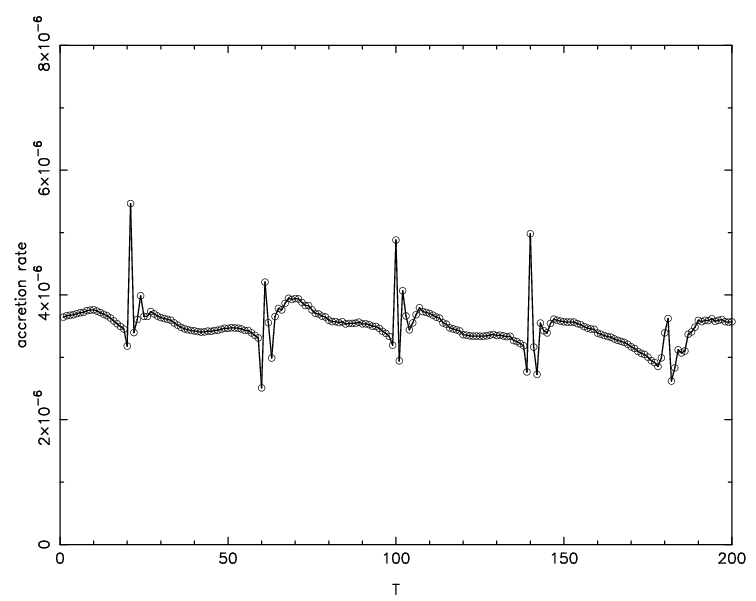}

    \caption{Same as Figure.~\ref{fig:figure2} but for D1 (up panel) and D2 (bottom panel).}
    \label{fig:figure5}
\end{figure}

Figure.~\ref{fig:figure1} shows the surface density of the accretion disk after about 400 orbits of the secondary black hole around the primary with eccentricity of 0.6 (Simulation A1) and 0.7 (Simulation A2) at apastron and periastron, with gap clearly seen. Figure.~\ref{fig:figure2} shows the accretion rate at the inner boundary which can be considered as the accretion rate onto the primary black hole approximately. We can clearly see from the up panel in Figure.~\ref{fig:figure2} that the accretion rate decreases as the secondary black hole with eccentricity of 0.6 (Simulation A1) approaches the periastron and reaches the minimum, then increases after it leaves the periastron. Unlike the up panel, the bottom panel of Figure.~\ref{fig:figure2} shows that the accretion rate tends to increase as the secondary black hole with eccentricity of 0.7 (Simulation A2) approaches the periastron and reaches the maximum at periastron, then decreases as the secondary black hole is far away from periastron. To show the effect of the mass of the secondary on the accretion onto the primary, Figure.~\ref{fig:figure3} shows the accretion rate at the inner boundary with the parameters in B1 and B2. The bottom panel in Figure.~\ref{fig:figure3} shows the similar property as that of the bottom panel in Figure.~\ref{fig:figure2} while the up panel of Figure.~\ref{fig:figure3} shows no periodicity. Results shown in Figure.~\ref{fig:figure4} from Simulation C1 shows no periodic accretion rate onto the primary SMBH in contrast to the results from Simulation A2 and B2. To illustrate the effect of semimajor of the secondary on the accretion onto the primary, Figure.~\ref{fig:figure5} presents the results from Simulation D1 and D2. With the periastron and mass of the primary and the secondary set to be the same as that of Simulation A2 and B2, Simulation D1 and D2 differ from A2 and B2 in that the semimajor in D1 and D2 is twice that in A2 and B2. As shown in Figure.~\ref{fig:figure5} periodicity of accretion rate tend to exist in SMBBH with large mass of the secondary, similar to property of studies in \citep{Hayasaki08,Sesana12} where the mass ratio of SMBBH is near unity and eccentricity is relatively small.

The results from the simulations demonstrate that the secondary black hole from the SMBBH can enhance the accretion rate onto the primary black hole with larger eccentricity and mass ratio than that with relatively small ones. SMBBH with intermediate mass ratio can have large eccentricity through some eccentricity exciting process, thus, there may exist many such SMBBH in the universe. Luminosity of quasar originates from the accretion onto the SMBH. Results from simulations in this work may give the observational prediction of periodic variability of the luminosity of quasar where there may exist such SMBBH adopted in this work. The bottom panel in Figure.~\ref{fig:figure2}, Figure.~\ref{fig:figure3} and Figure.~\ref{fig:figure5} show that the accretion rate can reach the maximum at periastron (luminosity is maximum at the same time), then, with the fact that gravitational radiation is maximum at periastron, the luminosity can be used as electromagnetic counterpart to the gravitational radiation in such system. Furthermore, simulation with large eccentricity may provide a possible mechanism to describe the potential periodic changing-look quasar since disk around the secondary black hole with extreme large eccentricity may be tidally disrupted and accreted by the primary black hole when reaching
the periastron. When the mass of secondary black hole is extremely small compared with that of the primary one, effect of the secondary black hole on the accretion onto the primary can be neglected. In this work, the accretion rate onto the secondary black hole is not investigated due to the fact that the number of the mesh should be largely increased in order to study the accretion rate onto the eccentric secondary black hole with small mass, which is largely time-expensive. We may expect from the Simulation D2 that the accretion rate onto the secondary black hole may decrease due to the tidal force from the primary SMBH which may tidally disrupt and accrete the disk surround the secondary when at periastron, and the accretion rate onto the secondary may increase by accreting from the disk when approaching apastron. We will investigate such situation in future.

\section{Conclusions}
We performed hydrodynamical simulations of tidal interaction between eccentric supermassive binary black hole with intermediate mass ratio and the surrounding gaseous disk in which the SMBBH is formed after the merger of two galaxies and excited to highly eccentric orbit due to the gravitational interaction between the SMBBH and the filed stars in the merged galaxy.

We mainly focus on the accretion onto the primary SMBH represented by the accretion rate onto the inner boundary of the simulated region. As shown by the 2D hydrodynamical simulation, the accretion rate onto the primary SMBH is dependent of the eccentricity, mass ratio and semimajor of the SMBBH obviously. Periodic accretion rate onto the primary SMBH tends to occur in SMBBH with small value of semimajor, large eccentricity and large mass ratio. Especially, accretion rate onto the primary SMBH can be strongly enhanced by the secondary with large eccentricity compared with the ones with small eccentricity under the same semimajor, and the secondary SMBH with large eccentricity can exert much more influence on the accretion of the primary SMBH than that with small eccentricity under the same periastron. Periodic accretion onto the primary SMBH may result in periodic luminosity which may be used as electromagnetic counterpart to gravitational wave radiated by such binary system. For high eccentricity such that $1-e \ll 1$, the pericenter distance can be roughly constant along the evolution dominated by gravitational radiation, thus, initial parameters adopted by this work could save much time of hydrodynamical simulation. For the secondary with extremely small mass, gravitational effect on the disk by the secondary black hole can be neglectable, which results that accretion onto the primary could be independent of the presence of the secondary.

With the limited number of simulations and the initial conditions parameter space, one cannot draw a too strong conclusion. But at the least, these simulations illustrate a range of outcomes. To describe the dynamics of this work more precisely, relativistic effect should be considered, such as the post-Newtonian metric of the binary felt by the fluid element with relativistic velocity and orbital evolution of the binary by gravitational radiation reaction. In this work, we conduct binary-disk interaction with Newtonian hydrodynamics. Relativistic simulations of interaction between eccentric SMBBH with intermediate mass ratio and disk will be performed in future.



\bsp	
\label{lastpage}
\end{document}